\documentclass{llncs}

\newcommand{\dw}[1]{\downarrow\!\!#1}
\newcommand{\bin}{\hbox{
\begin{picture}(1,1)(1,1)
\put(-0.5,1){$\times$}
\put(0,0){$\cup$}
\end{picture}
}\ }

\newcommand{\dwu}{\hbox{
\begin{picture}(1,1)(1,1)
\put(1,3){$\downarrow$}
\put(0,0){$\cup$}
\end{picture}
}\!\ }
\newcommand{\nopowersets}{SH^{\mbox{\sl w}}}
\newcommand{\nopow}{\mbox{\sl w}}
\newcommand{\nrel}{\overline{rel}}

\title{A study of set-sharing analysis via cliques}
\author{Jorge Navas$^{1}$ \ \ \ \ Francisco Bueno$^{2}$ \ \ \ \
        Manuel Hermenegildo$^{1,2}$ \\[0.05in]
        \texttt{jorge@cs.unm.edu, herme@unm.edu},\\
        \texttt{\{bueno, herme\}@fi.upm.es}}
\institute{
{\small $^1$Depts. of Comp. Science and Electr. and Comp. Eng.,
        Univ. of New Mexico, Albuquerque, NM, USA.}\\
{\small $^2$School of Computer Science, T. U. Madrid (UPM), Madrid, Spain}
}

\begin{document}
\maketitle

\begin{abstract}
  
  We study the problem of efficient, scalable set-sharing analysis of logic
  programs. We use the idea of representing sharing information as a pair
  of abstract substitutions, one of which is a worst-case sharing
  representation called a clique set, which was previously proposed for the
  case of inferring pair-sharing.  We use the clique-set representation for
  (1) inferring actual set-sharing information, and (2) analysis within a
  top-down framework. In particular, we define the abstract functions
  required by standard top-down analyses, both for sharing alone and also
  for the case of including freeness in addition to sharing. Our
  experimental evaluation supports the conclusion that, for inferring
  set-sharing, as it was the case for inferring pair-sharing, precision
  losses are limited, while useful efficiency gains are obtained. At the
  limit, the clique-set representation allowed analyzing some programs that
  exceeded memory capacity using classical sharing representations.

\end{abstract}

\section{Introduction}

In static analysis of logic programs the tracking of variables shared among
terms is essential. Arguably, the most accurate abstract domain defined for
tracking sharing is the Sharing domain~\cite{jacobs-jlp,ai-jlp}, which
represents variable occurrences, i.e., the possible occurrences of run-time
variables within the terms to which program variables will be bound. In
this paper we study an alternative representation for this domain.

\begin{example}
Let $V=\{x,y,z\}$ be a set of variables of interest. A substitution such
as $\{x/f(u_1,u_2,v_1,v_2,w),y/g(v_1,v_2,w),z/g(w,w)\}$ will be
abstracted in Sharing as $\{x,xy,xyz\}$.\footnote{To simplify
  notation, we denote a sharing group 
  (a set of variables representing sharing)  
  by the concatenation of its variables, e.g., $xyz$ is 
  $\{x,y,z\}$.} 
Sharing group $x$ in the abstraction represents the occurrence of
run-time variables $u_1$ and $u_2$ in the concrete substitution, $xy$
represents $v_1$ and $v_2$, and $xyz$ represents $w$. Note that 
the number of (occurrences of) run-time variables shared is abstracted
away.
\end{example}

Sharing analysis has been used for inferring several interesting
properties of programs; most notably (but not only), variable
independence. Several program variables are said to be independent if
the terms they are bound to do not have (run-time) variables in
common. Variable independence is the counterpart of sharing: program
variables share when the terms they are bound to do have run-time
variables in common. When we are talking of only two variables then we
refer to pair-sharing, and when it is more than two variables we refer
to set-sharing.
Sharing abstract domains are used to infer {\em possible} sharing,
i.e., the possibility that shared variables exist, and thus, in the
absence of such possibility, {\em definite} information about
independence.

\begin{example}
Let $V=\{x,y,z\}$ be variables of interest. A Sharing abstract substitution
such as $\{x,y,z\}$ (which denotes the set of the singleton sets
containing each variable) represents that all three variables are
independent. 
\end{example}

The Sharing domain has deserved a lot of attention in the literature
in the past. It has been enhanced in several
ways~\cite{file-sharing-tr,ZaffanellaBH99}. It has also 
been extended with other kinds of information, the most relevant of
which being
freeness and linearity~\cite{jacobs-jlp,codish96freeness,HillZB04TPLP},
but also for example information about term
structure~\cite{pathsh-iclp94,bruynooghe:composite,mulkers-iclp95}. 
Its combination with other abstract domains has also been studied to a
great extent~\cite{comdom,Fecht-plilp96}.
In particular, in~\cite{ZaffanellaBH99} an alternative representation for
Sharing is proposed for the non-redundant domain of~\cite{BagnaraHZ97SAS}
and this representation is thoroughly studied for inferring pair-sharing. A
new component is added to abstract substitutions that represents sets of
variables, the powerset of which would have been part of the original
abstract substitution. Such sets are called {\em cliques}.

\begin{example}
Let $V$ be as above. Consider the abstraction $\{x,xy,xyz,xz,y,yz,z\}$,
i.e., the powerset of $V$ (without the empty set). Such an abstraction
conveys no information: there might be run-time variables
shared by any pair of the three program variables, by the three of
them, or not shared at all. However, abstractions such as this one are
expensive to process during analysis: they penalize efficiency for no
benefit at all. The clique that will convey the same information is
simply the set $V$.
\end{example}

A clique is thus a compact representation for a piece of sharing which
in fact does not convey any useful information. 
The resulting precision and efficiency results for the case of inferring
pair-sharing were reported in~\cite{ZaffanellaBH99}.
In~\cite{ZaffanellaPhD} cliques are incorporated to the original Sharing
domain, but precision and efficiency are again studied for the case of
inferring pair-sharing.
Here, we are interested in studying precision and efficiency for the
different case of inferring set-sharing. Another difference with previous
work is that we develop the analysis for a top-down analysis framework,
which requires the definition of additional abstract functions in the
domain. Such functions were not defined in the previous works cited, since
bottom-up analyses were used there.

The rest of the paper proceeds as follows. Notation and preliminaries
are presented in Section~\ref{sec:prelim}. Then Section~\ref{sec:cliques}
introduces the representation based on cliques and the clique-domains
for set-sharing and set-sharing with freeness. In
Section~\ref{sec:topdown} the required functions for top-down
analysis are defined. In Section~\ref{sec:detecting} we present an
algorithm for detecting cliques, and in Section~\ref{sec:experiments}
our experimental evaluation of the proposed analyses. Finally,
Section~\ref{sec:conclude} concludes.

\section{Preliminaries}
\label{sec:prelim}

Let $\wp(S)$ denote the powerset of set $S$, and
$\wp^0(S)$ denote the {\em proper powerset} of set $S$, i.e.,
$\wp^0(S)=\wp(S)\setminus\{\emptyset\}$.
Let also $|S|$ denote the cardinality of a set $S$.

Let $V$ be a set of variables of interest; e.g., the variables of a
program.
A {\em sharing group} is a set of variables of interest, which
represents the possible sharing among them (i.e., that they might be
bound to terms which have a common variable). 
Let $SG=\wp^0(V)$ be the set of
all sharing groups. A {\em sharing set} is a set of sharing groups.
The Sharing domain is $SH=\wp(SG)$, the set of all sharing sets. 

For two elements $s_1\in SH$, $s_2\in SH$, 
let $s_1\bin s_2$ be their {\em binary union}, i.e., the
result of applying union to each pair in their Cartesian product
$s_1\times s_2$. 
Let also $s_1^*$ be the 
{\em star union} of $s_1$, i.e., its closure under union.
Given terms $s$ and $t$, and $sh\in SH$, we denote by
$sh_t$ the set of sets in $sh$ which have non-empty
intersection with the set of variables of $t$. By extension, in 
$sh_{st}$ $st$ acts as a single term. Also, $\overline{sh_{t}}$
is the complement of $sh_t$, i.e., $sh\setminus sh_t$.

Let $F$ and $P$ be sets of ranked (i.e., with a given arity)
functors of interest; e.g., the function symbols and the predicate
symbols of a program. We will use $Term$ to denote the set of terms
constructed from $V$ and $F\cup P$. Although somehow unorthodox, this
will allow us to simply write $g\in Term$ whether $g$ is a term or a
predicate atom, since all our operations apply equally well to both
classes of syntactic objects.
We will denote $\hat{t}$ the set of variables of $t\in Term$. 
For two elements $s\in Term$ and $t\in Term$, 
$\hat{st}=\hat{s}\cup\hat{t}$.

Analysis of a program proceeds by abstractly solving unification
equations of the form $t_1=t_2$, $t_1\in Term$, $t_2\in Term$.
Let $solve(t_1=t_2)$ denote the solved form of unification equation
$t_1=t_2$.
The results of analysis are abstract substitutions which approximate
the concrete substitutions that may occur during execution of the
program.
Let $U$ be a denumerable set of variables (e.g., the variables that
may occur during execution of a program). Concrete substitutions that
occur during execution are mappings from $V$ to the set of terms
constructed from $U\cup V$ and $F$.
Abstract substitutions are sharing sets.

\section{Clique domains}
\label{sec:cliques}

When a sharing set $sh\in SH$ includes the proper powerset of some set
$C$ of variables, the representation can be made more compact by using $C$ 
to represent the same sharing that its powerset represents in
the sharing set $sh$~\cite{ZaffanellaBH99}. The proper powerset of $C$
can then be eliminated from $sh$, since it is already represented by
$C$.
In fact, we will be using pairs $(cl,sh)$ of two sharing sets. The
second one represents sharing as in $SH$. However, in the first one,
each element $C\in cl$ represents the sharing that in $SH$ would be
represented by $\wp^0(C)$. 

A {\em clique} is, thus, a set of variables of
interest, much the same as a sharing group, but a clique $C$
represents all the sharing groups in $\wp^0(C)$. For a clique $C$, we
will use $\dw{C}=\wp^0(C)$. Note that $\dw{C}$ denotes all the
sharing that is implicitly represented in a clique $C$. 
A {\em clique set} is a set of cliques.
Let $CL=SH$ denote the set of all clique sets. 
For a clique
set $cl\in CL$ we define $\dwu{cl}=\cup\!\{\dw{C} ~|~ C\in cl\}$.
Note that $\dwu{cl}$ denotes all the
sharing that is implicitly represented in a clique set $cl$. 
For a pair $(cl,sh)$ of a clique set $cl$ and a sharing set $sh$, 
the sharing that the pair represents is $\dwu{cl}\cup sh$.

The Clique-Sharing domain is 
$\nopowersets=\{(cl,sh) ~|~ cl\in CL, sh\in SH\}$, i.e., the set of
pairs of a clique set and a sharing set~\cite{ZaffanellaBH99}.
An abstract unification operation $amgu^{\nopow}$ is defined
in~\cite{ZaffanellaPhD} which uses a 
function $\nrel:\wp(V)\times CL\longrightarrow CL$, 
defined as: 
\[\nrel(S,cl) = \{\ C\setminus S ~|~ C\in cl\ \} \setminus \{ \emptyset \} \]
and ($amgu^{\nopow}$) is equivalent to the following definition:

\[ amgu^s(x=t,(cl,sh))=\left\{
  \begin{array}{ll}
    (\ cl \ , \ \overline{sh_{xt}}\cup (sh_x^*\bin sh_t^*) \ )
  & \mbox{ if } cl_x=cl_t=\emptyset \\
    (\ \nrel(\hat{xt},cl) \ , \ \overline{sh_{xt}} \ )
  & \mbox{ if } cl_x=sh_x=\emptyset \\
  & \mbox{ or } cl_t=sh_t=\emptyset \\
    \multicolumn{2}{l}{ (\ \nrel(\hat{xt},cl)\cup 
      \{\cup(cl_x\cup cl_t\cup sh_x\cup sh_t)\} }\\
    \ , \ \overline{sh_{xt}} \ )  & \mbox{ otherwise }  \\
  \end{array}
  \right.
\]

Freeness can be introduced to the Clique-Sharing domain in the usual
way~\cite{freeness-iclp91}, by including a component which tracks the variables which are
known to be free. The Clique-Sharing+Freeness domain is thus 
$SHF^{\nopow} = \nopowersets\times V$.

Abstract unification $amgu^{sf}$ for equation $x=t$, $x\in V$, 
$t\in Term$, and $s\in SHF^{\nopow}$, $s=((cl,sh),f)$, is given
by $amgu^{sf}(x=t,s)=((cl',sh'),f')$, with:

\[(cl',sh')=\left\{
  \begin{array}{ll}
  amgu^{sff}(x=t,(cl,sh)) & \mbox{ if } x\in f \mbox{ or } t\in f \\
  amgu^{sfl}(x=t,(cl,sh)) & \mbox{ if } x\not\in f,\ t\not\in f 
                            \mbox{ and } lin^s(t) \\
  amgu^s(x=t,(cl,sh))     & \mbox{ otherwise } \\
  \end{array}
  \right.
\]
where $lin^s(t)$ holds iff $t$ is a linear term 
and\footnote{Note that checking this second condition can be rather
  expensive. Instead, the following, which is more efficient, can be
  checked: for all $s\in (sh_t\cup cl_t)$, $|s\cap\hat{t}|=1$.}
for all $\{y,z\}\subseteq\hat{t}$ such that $y\neq z$, 
$sh_y\cap sh_z=\emptyset$ and $cl_y\cap cl_z=\emptyset$; and:

\[\begin{array}{llll}
  amgu^{sff}(x=t,(cl,sh))=
  & ( & \nrel(\hat{xt},cl)\cup \\
  &   & ((cl_x\cup sh_x) \bin cl_t)\cup (cl_x \bin sh_t) \\
  & , & \overline{sh_{xt}}\cup (sh_x \bin sh_t ) & )
\end{array}
\]
\[amgu^{sfl}(x=t,(cl,sh)) = \left\{
  \begin{array}{ll}
  ( \ \nrel(\hat{xt},cl)\cup (cl_x \bin \{ \cup sh_t \} ) \\
  , \ \overline{sh_{xt}}\cup (sh_x \bin sh_t^* ) \ ) 
  & \mbox{ if } cl_t=\emptyset \\
  \multicolumn{2}{l}{  
  ( \ \nrel(\hat{xt},cl)\cup ((cl_x\cup sh_x) 
                        \bin \{ \cup(cl_t\cup sh_t) \}) } \\
  , \ \overline{sh_{xt}} \ )
  & \mbox{ if } cl_t\neq\emptyset
  \end{array}
  \right.
\]
\[ f'=\left\{
  \begin{array}{ll}
  f                                    & \mbox{ if } x\in f, t\in f \\
  f\setminus (\cup (sh_x \cup cl_x ))  & \mbox{ if } x\in f, t\not\in f \\
  f\setminus (\cup (sh_t \cup cl_t ))  & \mbox{ if } x\not\in f, t\in f \\
  f\setminus (\cup ((sh_x \cup cl_x ) \cup (sh_t \cup cl_t ))) 
                                       & \mbox{ if } x\not\in f, t\not\in f \\
  \end{array}
  \right.
\]

The operation $amgu^{sf}$ defined above is a simplification of the
corresponding operation which results from the method outlined
in~\cite{ZaffanellaPhD} to obtain an abstract unification for
$SH^{\nopow}$ plus freeness and linearity.

\section{Abstract functions required by top-down analysis}
\label{sec:topdown}

In top-down analysis frameworks, the analysis of a clause 
$Head${\tt :-}$Body$ is as follows. 
There is a goal $Goal$ for the predicate of $Head$, which is called in
a context represented by abstract substitution $Call$ on a set of 
variables (distinct from $\hat{Head}\cup\hat{Body}$)
which contains the variables of $Goal$. Then the success of $Goal$ by
executing the above clause is represented by abstract substitution 
$Succ$ given by:

\[\begin{array}{lcl}
Succ  & = & extend( Call, Goal, Prime ) \\
Prime & = & exit2succ( project( Head, Exit ), Goal, Head ) \\
Exit  & = & entry2exit( Body, Entry ) \\
Entry & = & augment( F, call2entry( Proj, Goal, Head ) ) \\
Proj  & = & project( Goal, Call ) 
\end{array}
\]
where $F$ is any term with the variables $\hat{Body}\setminus\hat{Head}$.
Function $project$ approximates the projection of a substitution on the
variables of a given term. Function $augment$ extends the domain of an
abstract substitution to the variables of a given term, which are
assumed to be new fresh variables. The rest of the
functions are as follows:

\begin{description}
\item[$call2entry(Proj,Goal,Head)$] ~\\
yields a substitution on the variables of
$Head$ which represents the effects of unification $Goal=Head$ in a
context represented by substitution $Proj$ on the variables of $Goal$.
\\

\item[$entry2exit(Body,Entry)$] ~\\
yields a substitution which represents the
success of $Body$ when called in a context represented by substitution
$Entry$. Both substitutions have a domain which includes the variables
of $Body$, and the domain of the resulting substitution includes the
domain of $Entry$.
\\

\item[$exit2succ(Exit',Goal,Head)$] ~\\
yields a substitution on the variables of
$Goal$ which represents the effects of unification $Goal=Head$ in a
context represented by substitution $Exit'$ on the variables of $Head$.
\\

\item[$extend(Call,Goal,Prime)$] ~\\
yields a substitution for the success of
$Goal$ when it is called in a context represented by substitution
$Call$ on a set of variables which contains the variables of $Goal$,
given that in such context the success of $Goal$ is already represented
by substitution $Prime$ on the variables of $Goal$. The domain of the
resulting substitution is the same as the domain of $Call$.
\end{description}

Function $entry2exit$ is given by the framework, and basically
traverses the body of a clause, analyzing each atom in turn.
The three domain-dependent abstract functions which are essential are:
$call2entry$, $exit2succ$, and $extend$. The first two can be defined
from the abstract unification operation $amgu$. The third 
one, however, is specific to the top-down framework and needs to be
defined specifically for a given domain.

Given an operation $amgu(x=t,ASub)$ of abstract unification for
equation $x=t$, $x\in V$, $t\in Term$, and $ASub$ an abstract 
substitution (the domain of which contains variables
$\hat{t}\cup\{x\}$), abstract unification for equation $t_1=t_2$,
$t_1\in Term$, $t_2\in Term$, is given by:

\[ unify( ASub, t_1, t_2) =
     project( t_1, Amgu( solve(t_1=t_2), augment(t_1,ASub) ) )
\]
\[ Amgu( Eq, ASub )=\left\{
   \begin{array}{ll}
   ASub                & \mbox{ if } Eq=\emptyset \\
   Amgu( Eq', amgu(x=t,ASub) ) & \mbox{ if } Eq=Eq'\cup\{x=t\}
   \end{array}
   \right.
\]
Functions $call2entry$ and $exit2succ$ can defined as
follows:
\[ \begin{array}{lcl}
    call2entry( ASub, Goal, Head) & = & unify( ASub, Head, Goal) \\
    exit2succ( ASub, Goal, Head)  & = & unify( ASub, Goal, Head) \\
   \end{array}
\]

However, $extend$, together with $project$, $augment$, and $amgu$ are
all domain-dependent.
In the Sharing domain, $extend$~\cite{ai-jlp}, $project$, and
$augment$ are defined as follows:

\[ extend(Call,g,Prime)=
   \overline{Call_{g}}\cup 
   \{\ s ~|~ s\in Call_{g}^*,\ (s\cap \hat{g})\in Prime\ \}
\]
\[ project(g,sh)= 
   \{ s \cap \hat{g} ~|~ s\in sh \}  \setminus \{ \emptyset \} 
\]
\[ augment(g,sh)= 
   sh\cup \{ \{x\} ~|~ x\in \hat{g} \}
\]
In the Sharing+Freeness domain, these functions are defined
as follows~\cite{freeness-iclp91}: 

\[ project^f(g,(sh,f)) = (project(g,sh),f\cap\hat{g})
\]
\[ augment^f(g,(sh,f)) = (augment(g,sh),f\cup\hat{g})
\]
\[ extend^f((sh_1,f_1),g,(sh_2,f_2)) = (sh',f') \]
\[ sh' = extend(sh_1,g,sh_2) \]
\[ f' = f_2\cup \{ x ~|~ x\in (f_1\setminus \hat{g}), 
                       ((\cup sh'_x)\cap\hat{g})\subseteq f_2 \}
\]

\subsection{Abstract functions for top-down analysis in the Clique-Domains}
\label{sec:functions}

Functions $call2entry$ and $exit2succ$ have usually been defined in a way
which is specific to the domain (see, e.g.,~\cite{ai-jlp} for a definition
for set-sharing). We have chosen instead to present here a formalization of
a way to use $amgu$ in top-down frameworks. Thus, the definitions of
$call2entry$ and $exit2succ$ based on $amgu$ given above. Our intuition in
doing this is that the results should be (more) comparable to
goal-dependent bottom-up analyses, where $amgu$ is used directly.

Note, however, that such definitions imply a possible loss of precision.
Using $amgu$ in the way explained above does not allow to take advantage of
the fact that all variables in the head of the clause being entered during
analysis are free. Alternative definitions of $call2entry$ can be obtained
that improve precision from this observation. The overall effect would be
equivalent to using the $amgu$ function for the Sharing domain coupled with
freeness, with the head variables as free variables, and then throwing out
the freeness component of the result. For example, for the Clique-Sharing
domain a function $call2entry^s$ can be defined as follows, where
$unify^{sf}$ is the version of $unify$ that uses $amgu^{sf}$:
\[ \begin{array}{lrcl}
   \multicolumn{2}{l}{
    call2entry^s( ASub, Goal, Head) } & = & ASub' \\
   \mbox{where} & ( ASub', Free ) & = & unify^{sf}( (ASub,\emptyset), Head, Goal)
   \end{array}
\]

However, for the reasons mentioned above, we have used the definitions
of $call2entry$ and $exit2succ$ based on $amgu$.
The rest of the top-down functions are defined below. For the
Clique-Sharing domain, let $g\in Term$, and $(cl,sh)\in\nopowersets$.
Functions $project^s$ and $augment^s$ are defined as follows:
\[ project^s(g,(cl,sh))=(project(g,cl),project(g,sh)) \]
\[ augment^s(g,(cl,sh)) = (cl,augment(g,sh)) \]
Function $extend^s(Call,g,Prime)$ is defined as follows.
Let $Call=(cl_1,sh_1)$ and $Prime=(cl_2,sh_2)$. Let $normalize$ be a
function which normalizes a pair $(cl,sh)$ so that no powersets occur
in $sh$ (all are ``transferred'' to cliques in $cl$;
Section~\ref{sec:detecting} presents a possible implementation of such
a function). 
Let $Prime$ be already normalized, and:
\[ (cl',sh') = 
   normalize(({cl_1}_{g}^*\cup ({cl_1}_{g}^* \bin {sh_1}_{g}^*),{sh_1}_{g}^*))
\]

The following two functions lift the classical $extend$~\cite{ai-jlp}
respectively to the cases of the two clique sets and of the two
sharing sets occurring in each of the pairs in $Call$ and $Prime$: 
\[extsh(sh_1,g,sh_2)=
  \overline{{sh_1}_g}\cup 
  \{\ s ~|~ s\in sh',\ (s\cap \hat{g})\in sh_2\ \}
\]
\[extcl(cl_1,g,cl_2)=
  \overline{rel}(\hat{g},cl_1)\cup 
  \{\ (s'\cap s) \cup (s'\setminus\hat{g}) ~|~ s'\in cl',\ s\in cl_2\ \}     
\]

The following two functions account respectively for the cases of the
clique set of $Call$ and the sharing set of $Prime$, and the other way
around:
\[ clsh(cl',g,sh_2)=\{\ s ~|~ s\subseteq c\in cl',\ (s\cap \hat{g})\in sh_2\ \}
\]
\[ shcl(sh',g,cl_2)=\{\ s ~|~ s\in sh',\ (s\cap \hat{g})\subseteq c\in cl_2\ \}
\]

The function extend for the Clique-Sharing domain is thus:
\[\begin{array}{lll}
  \multicolumn{2}{l}{
  extend^s((cl_1,sh_1),g,(cl_2,sh_2)) \ = } \\
  ( & extcl(cl_1,g,cl_2) \\
  , & extsh(sh_1,g,sh_2) \cup clsh(cl',g,sh_2) \cup shcl(sh',g,cl_2)
  & )
\end{array}\]

\begin{example}
Let $Call=(cl_1,sh_1)=(\{xyz\},\{u,v\})$, $Prime=(cl_2,sh_2)= (\{x\},\{uv\})$,
and $\hat{g}=\{x,u,v\}$. Then we have $(cl',sh')=(\{xyzuv\},\emptyset)$. The
function $extend^s$ is computed as follows:
\[ extsh(sh_1,g,sh_2) = extsh(\{u,v\},g,\{uv\}) = \emptyset \]
\[ extcl(cl_1,g,cl_2) = extcl(\{xyz\},g,\{x\}) = \{xyz,yz\} \]
\[ clsh(cl',g,sh_2) =  clsh(\{xyzuv\},g,\{uv\}) = \{yzuv,yuv,zuv,uv\} \]
\[ shcl(sh',g,cl_2) =  shcl(\emptyset,g,\{x\}) = \emptyset \]
Thus, $extend^s(Call,g,Prime)= (\{xyz,yz\},\{yzuv,yuv,zuv,uv\})$,
which after regularization yields $(\{xyz\},\{yzuv,yuv,zuv,uv\})$.
\end{example}

Note how the result is less precise than the exact result
$(\{xyz\},\{uv\})$. This is due to overestimation of sharing implied
by the cliques; in particular, for the case of $extend$,
overestimations stem mainly from the necessary worst-case assumption
given by $(cl',sh')$, which is then ``pruned'' as much as possible by
the functions defined above.

\begin{theorem}
Let $Call\in\nopowersets$, $Prime\in\nopowersets$, and
$g\in Term$, such that the conditions for the $extend$ function are
satisfied. Let $Call=(cl_1,sh_1)$, $Prime=(cl_2,sh_2)$,
and $extend^s(Call,g,Prime)=(cl',sh')$. Then
\[ (\dwu{cl'}\cup sh') \supseteq 
   extend(\dwu{cl_1}\cup sh_1,g,\dwu{cl_2}\cup sh_2) \ .
\]
\end{theorem}

For the Clique-Sharing+Freeness domain, 
let $g\in Term$, and $s\in SHF^{\nopow}$, $s=((cl,sh),f)$.
Functions $project^{sf}$ and $augment^{sf}$ are defined as follows:

\[ project^{sf}(g,s) = (project^s(g,(cl,sh)),f\cap\hat{g}) \]
\[ augment^{sf}(g,s) = (augment^s(g,(cl,sh)),f\cup\hat{g}) \]
Function $extend^{sf}(Call,g,Prime)$ is defined as follows.
Let $Call=((cl_1,sh_1),f_1)$ and $Prime=((cl_2,sh_2),f_2)$,
$extend^{sf}(Call,g,Prime)=((cl',sh'),f')$, where:

\[ (cl',sh')=extend^s((cl_1,sh_1),g,(cl_2,sh_2)) \]
\[ f'=f_2\cup \{ x ~|~ x\in (f_1\setminus \hat{g}),\ 
                    ((\cup (sh'_x \cup cl'_x))\cap\hat{g})\subseteq f_2 \}
\]

\begin{theorem}
Let $Call\in SHF^{\nopow}$, $Prime\in SHF^{\nopow}$, and
$g\in Term$, such that the conditions for the $extend$ function are
satisfied. Let $Call=((cl_1,sh_1),f_1)$, $Prime=((cl_2,sh_2),f_2)$,
and $extend^{sf}(Call,g,Prime)=((cl',sh'),f')$.
Let also $s_1=\dwu{cl_1}\cup sh_1$, $s_2=\dwu{cl_2}\cup sh_2$,
and $extend^f((s_1,f_1),g,(s_2,f_2))=(sh,f)$. Then
\[ (\dwu{cl'}\cup sh') \supseteq sh \ \mbox{ and }\ f' \subseteq f \ . \]
\end{theorem}

\section{Detecting cliques}
\label{sec:detecting}

Obviously, to minimize the representation in $\nopowersets$ it pays
off to replace any set $S$ of sharing groups which is the proper
powerset of some set of variables $C$ by including $C$ as a
clique. Once this is done, the set $S$ can be eliminated from the
sharing set, since the presence of $C$ in the clique set makes $S$
redundant. 
This is the normalization mentioned in Section~\ref{sec:functions}
when defining $extend$ for the Clique-Sharing domain, and denoted
there by a function $normalize$.
In this section we present an algorithm for such a normalization.

Given an element $(cl,sh)\in\nopowersets$, sharing groups might occur
in $sh$ which are already implicit in $cl$. Such groups are redundant
with respect to the sharing represented by the pair. 
We say that an element $(cl,sh)\in\nopowersets$ is \emph{minimal} if
$\dwu{cl}\cap sh=\emptyset$.
An algorithm for minimization is straightforward: it should delete
from $sh$ all sharing groups which are a subset of an existing clique
in $cl$. 
%
But normalization goes a step further by ``moving sharing'' from the
sharing set of a pair to the clique set, thus forcing redundancy of
some sharing groups (which can therefore be eliminated).

While normalizing, it turns out that powersets may exist which can be
obtained from sharing groups in the sharing set plus sharing groups
implied by existing cliques in the clique set. The representation can
be minimized further if such sharing groups are also ``transferred'' to
the clique set by adding the adequate clique.
We say that an element $(cl,sh)\in\nopowersets$ is \emph{normalized} if
whenever there is an $s\subseteq (\dwu{cl}\cup sh)$ such that $s=\dw{c}$
for some set $c$ then $s\cap sh=\emptyset$.

It is important to stress the fact that neither minimization nor
normalization change the precision of the sharing representation. They
are both {\em reductions}, or compressions of the representation of a
substitution, in the sense that the substitution is the same (i.e.,
conveys the same information) but its representation is smaller. Thus,
they are not a widening operation, in the sense, widely used, of a
change in domain or representation with the objective of improving
efficiency at the cost of losing precision. This is not the case in
the above operations.

Our normalization algorithm is presented in
Figure~\ref{fig:detecting}. It starts with an element
$(cl,sh)\in\nopowersets$, which is already minimal, and obtains an
equivalent element (w.r.t.\ the sharing represented) which is
normalized.
First, the number $m$ is computed, which is the length of the longest
possible clique. Then the sharing set $sh$ is traversed to obtain candidate
cliques of the greatest possible length $i$ (which starts in $m$ and
is iteratively decremented). Existing subsets of a candidate
clique $S$ of length $i$ are extracted from $sh$. If there are $2^i-1-[S]$
subsets of $S$ in $sh$ then $S$ is a clique: it is added to $cl$ and
its subsets deleted from $sh$. 
Note that the test is performed on the number of existing subsets,
and requires the computation of a number $[S]$, which is crucial
for the correctness of the test.

\begin{figure}
\centering
\begin{minipage}{0.6\textwidth}
\begin{enumerate}
\item Let $n=|sh|$; if $n < 3$, stop.
\item Compute the maximum $m$ such that $n\geq 2^m-1$.
\item Let $i = m$.
\item If $i = 1$, stop.
\item Let $C=\{ s ~|~ s\in sh, |s|=i \}$.
\item If $C=\emptyset$ then decrement $i$ and go to 4.
\item Take $S\in C$ and delete it from $C$.
\item Let $SS=\{ s ~|~ s\in sh, s\subseteq S \}$.
\item Compute $[S]$.
\item If $|SS|=2^i-1-[S]$ then:
  \begin{enumerate}
  \item Add $S$ to $cl$ (regularize $cl$).
  \item Subtract $SS$ from $sh$.
  \end{enumerate}
\item Go to 6.
\end{enumerate}
\end{minipage}
\caption{Algorithm for detecting cliques}
\label{fig:detecting}
\end{figure}

The number $[S]$ corresponds to the number of subsets of $S$ which
may not appear in $sh$ because they are already represented in $cl$
(i.e., they are already subsets of an existing clique). In order to
correctly compute this number it is essential that the input to the
algorithm is already minimal; otherwise, redundant sharing groups
might bias the calculation: the formula below may count as not present
in $sh$ a (redundant) group which is in fact present.
The computation of $[S]$ is as follows. 
Take $cl$ in its state at step 9 of the algorithm.
Let $I = \{ S\cap C ~|~ C \in cl \}\setminus \{\emptyset\}$ and 
$A_i = \{ \cap A ~|~ A \subseteq I,\ |A|=i \}$. Then:

\[ [S] = \sum_{1\leq i \leq |I|} (-1)^{i-1}
         \sum_{A \in A_i} (2^{|A|}-1)
\]

Note that the representation can be minimized further by eliminating
cliques which are redundant with other cliques. This is the
regularization mentioned in step 10 of the algorithm.
We say that a clique set $cl$ is \emph{regular} if there are no two
cliques $c_1\in cl$, $c_2\in cl$, such that $c_1\subset c_2$.
This can be tested while adding cliques in step 10 above.

Finally, there is a chance for further minimization by considering as
cliques candidate sets of variables such that not all of their subsets
exist in the given element of $\nopowersets$. This opens up the
possibility of using the above algorithm as a widening.
Note that the algorithm preserves precision, since the sharing
represented by the element of $\nopowersets$ input to the algorithm is
the same as that represented by the element which is output.
However, we could set up a threshold for the number of subsets of the
candidate clique that need be detected, and in this case the output
element may in general represent more sharing. 
%

\section{Experimental results}
\label{sec:experiments}

We have measured experimentally the relative efficiency and precision
obtained with the inclusion of cliques in the Sharing and Sharing+Freeness
domains. 
We measure absolute precision of a sharing set by the number of its
sharing groups relative to the number of sharing groups in the
worst-case for the set of variables in its domain.
The number of sharing groups in the worst-case sharing for $n$
variables is given by $2^n-1$. 

Our results are shown in Tables~\ref{table:share} for Sharing
and~\ref{table:sharefree} for Sharing+Freeness. Columns labeled {\bf time}
show analysis times in milliseconds. on a medium-loaded Pentium IV Xeon
2.0Ghz with two processors, 4Gb of RAM memory, running Fedora Core 2.0, and
averaging several runs after eliminating the best and worst values. Ciao
version 1.11\#326 and CiaoPP 1.0\#2292 were used. Columns labeled {\bf
  precision} show the number of sharing groups in the information inferred
and, between parenthesis, the number of sharing groups for the worst-case
sharing. Columns labeled {\bf \#C} show the number of clique groups. In
both tables, first the numbers for the original domain are shown, then the
numbers for the clique-domain.
Since our analyses infer information at all program points (before and
after calling each clause body atom), and also several variants for each
program point, we show the accumulated number of sharing groups in all
variants for all program points.

\begin{table}[htbp]
\vspace{-\baselineskip}
  \centering
  \small
\begin{tabular}{|l||r|r|r||r|r|r|}
\hline
\multicolumn{1}{|c||}{} &\multicolumn{3}{c||}{\bf Sharing} 
 &\multicolumn{3}{c|}{\bf Clique-Sharing} \\
\cline{2-7}
   & \multicolumn{1}{|c}{\textbf{time}}
   & \multicolumn{1}{|c|}{\textbf{precision}}   
   & \multicolumn{1}{|c||}{\textbf{ \#C}}
   & \multicolumn{1}{|c}{\textbf{time}}
   & \multicolumn{1}{|c|}{\textbf{precision}}   
   & \multicolumn{1}{|c|}{\textbf{ \#C}} \\
\hline 
append & 11 & 29 (60) & 0 & 8 & 44 (60) & 4  \\ 
\hline 
deriv & 35 & 27 (546) & 0 & 27 & 27 (546) & 0  \\  
\hline 
mmatrix & 13 & 14 (694) & 0 & 11 & 14 (694) & 0  \\  
\hline 
qsort & 24 & 30 (1716) & 0 & 25 & 30 (1716) & 0  \\  
\hline 
query & 11 & 35 (501) & 0 & 13 & 35 (501) & 5  \\  
\hline 
serialize & 306 & 1734 (10531) & 0 & 90 & 2443 (10531) & 88  \\  
\hline
\hline  
aiakl & 35 & 145 (13238) & 0 & 42 & 145 (13238) & 0 \\  
\hline 
boyer & 369 & 1688 (4631) & 0 & 267 & 1997 (4631) & 158 \\  
\hline 
browse & 30 & 69 (776) & 0 & 29 & 69 (776) & 0  \\  
\hline 
prolog\_read & 400 & 1080 (408755) & 0 & 465 & 1080 (408755) & 10 \\  
\hline 
rdtok & 325 & 1350 (11513) & 0 & 344 & 1391 (11513) & 182 \\   
\hline 
warplan & 3261 & 8207 (42089) & 0 & 1430 & 8191 (26857) & 420  \\  
\hline 
zebra & 25 & 280 (671088746) & 0 & 34 & 280 (671088746) & 0 \\ 
\hline 
\hline 
ann & 2382 & 10000 (314354) & 0 & 802 & 19544 (313790) & 700  \\ 
\hline 
peephole & 831 & 2210 (12148) & 0 & 435 & 2920 (12118) & 171 \\  
\hline 
qplan    & -  & - & - & 860 & 420203 (3826458) & 747  \\
\hline 
witt & 405 & 858 (4545564) & 0 & 437 & 858 (4545564) & 25 \\  
\hline 

\end{tabular}
\caption{Precision and Time-efficiency for Sharing}
\label{table:share}
\vspace{-1\baselineskip}
\end{table}

Benchmarks are divided into three groups. 
Of each group we only show a reduced number of the benchmarks actually
used: those which are more representative.
The first group, append
through serialize, is a set of simple programs, used as a testbed for
an analysis: they have only direct recursion and make a
straightforward use of unification (basically, for input/output of
arguments). The second group, aiakl through zebra, are more involved:
they make use of mutual recursion and of elaborated aliasing between
arguments to some extent; some of them are parts of ``real'' programs
(aiakl is part of an analyzer of the AKL language; prolog\_read and
rdtok are parsers of Prolog). The benchmarks in the third group are
all (parts of) ``real'' programs: ann is the \&-prolog parallelizer,
peephole is the peephole optimizer of the SB-Prolog compiler, qplan
is the core of the Chat-80 application, and witt is a conceptual
clustering application. 

\begin{table}[htbp]
\vspace{-\baselineskip}
  \centering
  \small
\begin{tabular}{|l||r|r|r||r|r|r|}
\hline
\multicolumn{1}{|c||}{} &\multicolumn{3}{c||}{\bf Sharing+Freeness} 
 &\multicolumn{3}{c|}{\bf Clique-Sharing+Freeness} \\
\cline{2-7}
   & \multicolumn{1}{|c}{\textbf{time}}
   & \multicolumn{1}{|c|}{\textbf{precision}}   
   & \multicolumn{1}{|c||}{\textbf{ \#C}}
   & \multicolumn{1}{|c}{\textbf{time}}
   & \multicolumn{1}{|c|}{\textbf{precision}}   
   & \multicolumn{1}{|c|}{\textbf{ \#C}} \\
\hline 
append &  6 & 7 (30) & 0 & 6 & 7 (30) & 0  \\ 
\hline 
deriv  & 27 & 21 (546) & 0 & 27 & 21 (546) & 0  \\  
\hline 
mmatrix &  9 & 12 (694) & 0 & 11 & 12 (694) & 0  \\  
\hline 
qsort &  25 & 30 (1716) & 0 & 27 & 30 (1716) & 0  \\  
\hline 
query & 12 & 22 (501) & 0 & 14 & 22 (501) & 0  \\  
\hline 
serialize &  61 & 545 (5264) & 0 & 55 & 736 (5264) & 41  \\  
\hline
\hline  
aiakl &  37 & 145 (13238) & 0 & 43 & 145 (13238) & 0  \\  
\hline 
boyer &  373 & 1739 (5036) & 0 & 278 & 2074 (5036) & 163  \\  
\hline 
browse & 29 & 69 (776) & 0 & 31 & 69 (776) & 0  \\  
\hline 
prolog\_read &  425 & 1050 (408634) & 0 & 481 & 1050 (408634) & 0  \\  
\hline 
rdtok &  335 & 1047 (11513) & 0 & 357 & 1053 (11513) & 2  \\   
\hline 
warplan &  1320 & 3068 (23501) & 0 & 1264 & 5705 (25345) & 209  \\  
\hline 
zebra   & 41 & 280 (671088746) & 0 & 42 & 280 (671088746) & 0  \\ 
\hline 
\hline 
ann & 1791 & 7811 (401220) & 0 & 968 & 14108 (394800) & 510  \\ 
\hline 
peephole & 508 & 1475 (9941) & 0 & 403 & 2825 (12410) & 135  \\  
\hline 
qplan   &  - & - & - & 2181 & 233070 (3126973) & 529 \\
\hline 
witt & 484 & 813 (4545594) & 0 & 451 & 813 (4545594) & 0  \\  
\hline 
\end{tabular}
\caption{Precision and Time-efficiency for Sharing+Freeness}
\label{table:sharefree}
\vspace{-\baselineskip}
\end{table}

In order to understand the results shown in the tables above it is 
important to note an existing synergy between normalization,
efficiency, and precision. 
If normalization causes no change in the sharing representation (i.e.,
sharing groups are not moved to cliques), usually because powersets do not
really occur during analysis, then the clique part is empty.  Analysis is
the same as without cliques, but with the extra overhead due to the
use of the normalization process. Then precision is the same but the time
spent in analyzing the program is a little longer.
This also occurs often if the use of normalization is kept to a minimum:
only for correctness (in our implementation, normalization is required
for correctness at least for the $extend$ function and other
functions used for comparing abstract substitutions). This should not
be surprising, since the fact that powersets occur during analysis at
a given time does not necessarily mean that they keep on occurring
afterward: they can disappear because of groundness or other
precision improvements during subsequent analysis (of, e.g.,
builtins). 

When the normalization process is used more often (like for example at
every call to $call2entry$ as we have done), then sharing groups are moved
more often to cliques. Thus, the use of the operations that compute on
clique sets produces efficiency gains, and also precision losses, as it was
expected. However, precision losses are not high. Finally, if normalization
is used too often, then the analysis process suffers from heavy overhead,
causing too high penalty in efficiency.
Therefore it is very clear that a thorough tuning of the use of the
normalization process is crucial to lead analysis to good results in
terms of both precision and efficiency.  

As usual in top-down analysis, the $extend$ function plays a crucial role.
In our case, this function is a very important bottleneck for the use of
normalization. As we have said, we use the normalization for correctness at
the beginning of the function $extend$. Additionally, it would be
convenient to use it also at the end of such function, since the number of
sharing groups can grow too much. However, this is not possible due to the
$clsh$ function, which can generate so many sharing groups that, at the
limit, the normalization process itself cannot run. Alternative definitions
of $clsh$ have been studied, but because of the precision losses incurred,
they have been found impractical.

From the above tables we can notice that there are always programs the
analysis of which does not produce cliques. This shows up in some of the
benchmarks (like all of the first group but serialize and some of the
second one such as aiakl, browse, prolog\_read, and zebra).  In this case,
as it was expected, precision is maintained but there is a small loss of
efficiency due to the commented extra overhead.
The same thing happens with benchmarks which produce cliques, but this
does not affect precision: append, query, prolog\_read, and witt, in
the case of Sharing without freeness.

On the other hand, for those benchmarks which do generate cliques (like
serialize, boyer, warplan, ann, and peephole) the gain in efficiency is
considerable at the cost of a small precision loss.  As usual, efficiency
and precision correlate inversely: if precision increases then efficiency
decreases and vice versa.  A special case is, to some extent, that of
rdtok, since precision losses are not coupled with efficiency gains. The
reason is that for this benchmark there are extra success substitutions
(which do not convey extra precision and, in fact, the result is less
precise) that make the analysis runs longer.

In general, the same effects are maintained with the addition of freeness,
although the efficiency gains are lower whereas the precision gains are a
little higher.  The reason is that the function $amgu^{sf}$ is less
efficient than $amgu^{s}$ (but more precise).  Overall, however, the
trade-off between precision and efficiency is beneficial.  Moreover, the
more compact representation of the clique domain makes possible to analyze
benchmarks (e.g., qplan) which run out of memory with the standard
representation.

\paragraph{\bf{Effectiveness}.} We have also tested how relevant precision
losses can be when the analysis is used as part of another application. 
In particular, we have used the Clique-Sharing+Freeness domain for
inferring non-failure information~\cite{nfplai-flops04}. We have
selected a representative subset of our benchmarks. 
Results for them are shown in Table~\ref{table:non_failure}.
Columns marked {\bf Total} show the number of predicates. Columns marked
{\bf NF} show the number of predicates which the analysis can infer that
they will not fail. Columns marked {\bf Cov} show the number of predicates
that the analysis can infer that they are covered (a necessary condition
for guaranteeing non-failure).
The results obtained suggest that the precision losses caused by the use of
the clique domain are not relevant when the information from analysis is
used as input in this particular application.

\begin{table}[htbp]
\vspace{-\baselineskip}
  \centering
  \small
\begin{tabular}{|l||r|r|r||r|r|r|}
\hline
\multicolumn{1}{|c||}{} &\multicolumn{3}{c||}{\bf Sharing+Freeness} 
 &\multicolumn{3}{c|}{\bf Clique-Sharing+Freeness} \\
\cline{1-7}
   & \multicolumn{1}{|c}{\textbf{Total}}
   & \multicolumn{1}{|c|}{\textbf{NF (\%)}}   
   & \multicolumn{1}{|c||}{\textbf{Cov (\%)}}
   & \multicolumn{1}{|c}{\textbf{Total}}
   & \multicolumn{1}{|c|}{\textbf{NF (\%)}}   
   & \multicolumn{1}{|c|}{\textbf{Cov (\%)}} \\
\hline 
append & 1 & 1 (100) & 1 (100) & 1 & 1 (100) & 1 (100) \\ 
\hline 
deriv & 1 & 1 (100) & 1 (100) & 1 & 1 (100) & 1 (100) \\ 
\hline 
qsort  & 3 & 3 (100) & 3 (100) & 3 & 3 (100) & 3 (100) \\ 
\hline 
serialize & 5 & 0 (0) & 2 (40) & 5 & 0 (0) & 2 (40) \\ 
\hline 
\hline 
rdtok & 22 & 8 (36) & 13 (59) & 22 & 8 (36) & 13 (59) \\ 
\hline 
zebra & 6 & 1 (16)  & 4 (66) & 6 & 1 (16) & 4 (66) \\ 
\hline 
\end{tabular}
\caption{Accuracy of the non-failure analysis}
\label{table:non_failure}
\vspace{-\baselineskip}
\end{table}

\section{Conclusions and Future work}
\label{sec:conclude}

We have reported on a study of efficiency and precision of the clique
representation of sharing when used for inferring proper set-sharing,
as opposed to pair-sharing. We have also included the case of
Clique-Sharing plus freeness information. 
Besides the abstract unification operations for both domains with the
clique representation (equivalent definitions of which were already
proposed in the literature), we have contributed other operations required
for top-down analyses, in particular, the extend function. 
Experiments reported aim specifically at the use of cliques as an
alternative representation, not as a widening (as opposed to similar
experiments reported in~\cite{ZaffanellaPhD}, where a threshold on the
number of allowed sharing groups was imposed that triggered their move
into cliques). 
We are currently working on using the clique representation as a
widening in order to solve the mentioned limitations of the 
$extend$ function. 
In line with the conclusions from previous experiments, our
experimental evaluation also supports the conclusion that precision losses
are reasonable. This is also supported additionally by our experiments
in actually using the information inferred, as we have showed for
inferring non-failure. 
Efficiency gains have also been shown, to the extreme case of being able to
analyze programs that exceeded memory capacity using the classical sharing
representation.

\section*{Acknowledgements}

{\small The authors would like to thank the anonymous referees for their
  useful comments.  Manuel Hermenegildo and Jorge Navas are supported in
  part by the Prince of Asturias Chair in Information Science and
  Technology at UNM. This work was also funded in part by the EC Future and
  Emerging Technologies program IST-2001-38059 {\em ASAP} project and by
  the Spanish MEC TIC 2002-0055 {\em CUBICO} project. }

\begin{small}
\newcommand{\etalchar}[1]{$^{#1}$}

\end{small}

\end{document}